\documentclass{aa}

\usepackage{graphicx}
\usepackage{txfonts} \usepackage{natbib}

\begin{document}

   \title{Constraining physics of very hot super-Earths with the James Webb Telescope. The case of Corot-7b.}

   \author{B. Samuel
		  \inst{1}
          	   \and
               J. Leconte
		  \inst{2}
	         \and
               D. Rouan
		   \inst{1}
	         \and
               F. Forget
		   \inst{2}
	        \and
              A. L\'eger
	        \inst{3}
	        \and
               J. Schneider
	 	  \inst{4}
               }

   \institute{LESIA, UMR 8109 CNRS, Observatoire de Paris, UVSQ, Université Paris-Diderot, 5 place J. Janssen, 92195 Meudon, France
              \email{benjamin.samuel@obspm.fr}
         \and
          Laboratoire de M\'et\'eorologie Dynamique, Universit\'e Paris 6, Paris, France
          \and
          IAS, CNRS (UMR 8617), Universit\'e Paris-Sud, Orsay, France
          \and
            LUTH, Observatoire de Paris, CNRS, Universit\'e Paris Diderot ; 5 Place Jules
Janssen, 92190 Meudon, France  
             }
  \authorrunning{Samuel, Leconte, Rouan et al.} \date{Received 3 January 2013; accepted 18 December 2013}

% \abstract{}{}{}{}{} 5 {} token are mandatory
% 
  \abstract
  % context heading (optional) {} leave it empty if necessary  
   { Super-Earths with solid surfaces such CoRot-7b and Kepler-10b are expected to be extremely hot. It has been suggested that they must be atmosphere-free and that a lava ocean is present on their hot dayside. Here, we use several dedicated thermal models to study how observations with NIRSPEC on the JWST could further confirm and constrain, or reject the atmosphere-free lava ocean planet model for very hot super-Earths. Using CoRoT-7b as a working case, we explore the consequences on the phase-curve of a non tidal-locked rotation, with/without an atmosphere, and for different values of the albedo. We simulate future observations of the reflected light and thermal emission from CoRoT-7b with NIRSPEC-JWST and look for detectable signatures, such as time lag, as well as the possibility to retrieve the latitudinal surface temperature distribution.
We demonstrate that we should be able to constrain several parameters after observations of two orbits (42 h) with a broad range of wavelengths: i)The Bond albedo is retrieved to within +/- 0.03 in most cases. ii) The lag effect allows us to retrieve the rotation period within 3 hours for a planet, with rotation half the orbital period. iii) Any spin period shorter than a limit in the range 30 - 800 h, depending on the thickness of the thermal layer in the soil, would be detected. iv) The presence of a thick gray atmosphere with a pressure of one bar, and a specific opacity higher than $10^{-5} m^{-2} kg^{-1}$ is detectable. v) With spectra up to 4.5 $\mu$m, the latitudinal temperature profile can be retrieved to within 30 K with a risk of a totally wrong solution in 5 \% of the cases. We conclude that it should thus be possible to distinguish the reference situation from other cases. Surface temperature map and the albedo brings important constraints on the nature or the physical state of the soil of hot super-Earths. }

   \keywords{planets and satellites: surfaces -- planets and satellites:
   atmospheres -- planet-star interactions -- planets and satellites:
   physical evolution -- planets and satellites: fundamental parameters
   -- planets and satellites: composition
               }
               
   \maketitle

%
%________________________________________________________________
%
\section{Introduction}

During the last decade, the development of high precision photometry,
particularly from space, led to one of the most exciting, albeit
expected, accomplishment:  the first detection of massive rocky planets,
or the so-called super-Earths (SE). Despite their large masses, the
terrestrial nature of these super-Earths has been confirmed by the
measure of their mean densities by combining both transit and radial
velocity observations. Surprisingly, the first two objects that are
observed compatible with a rocky composition, CoRoT-7b
\citep{Leger_2009, Queloz_2009, Hatzes_2011}  and KEPLER-10b
\citep{Batalha_2011}, appear to be similar in radius ($1.68 \pm 0.09$
and $1.42 \pm 0.03  \, \mathrm{R_\oplus}$), mass ($7.42 \pm 1.21$  and
$4.56 \pm 1.2  \, \mathrm{M_\oplus}$), and environment (They both orbit
around a main sequence G star with a very short period roughly equal to
0.85 day). However, their formation is far from being understood. In
particular, whether they could have formed {\it in situ} or should have
migrated to this close orbit or whether they formed as rocky planets or
as gas giants whose gaseous envelope subsequently evaporated
\citep{Leitzinger_2011} remains unclear.

Considering the magnitude of the energy dissipation induced by the
unusually strong tides, this kind of rocky objects should most likely be
phase-locked in a very short timescale ($<1$ My). Furthermore, the
extreme UV fluxes and stellar winds received by the planet ensure an
efficient atmospheric escape. As a result, if an atmosphere is present,
it is probably very thin and in condensation/sublimation equilibrium
with the molten surface, although the maximum possible pressure of such
an atmosphere is still debated (\citealt{SF09}; \citealt{Leger_2011}
-hereafter L11-; \citealt{CM11}; \citealt{Kite_2011}).

According to L11, \citet{Mura_2011}, and \citet{Rouan_2011}, the direct
expected consequences on the planet surface properties are as follow:

\begin{itemize} \item A violent thermal gradient between the day/night
hemispheres (2500-3000 K versus 50 K) because of the absence of thermal
surface redistribution. \item A large lava ocean on the day side, where
the temperature exceeds the rock fusion temperature. \item A
cometary-like tail trailing the planet \citep{Mura_2011}. These tails
have already been detected for Mercury (e.g. \citealt{Potter_2002}) and
possibly for HD 209458 b  \citep{Vidal-Madjar_2003}. Depending on the
atmospheric molecular composition, this tail can be observable.
Constraints on the molecules present in the atmosphere of exoplanets
have already been inferred from observations with the VLT
\citep{Guenther_2011}. \end{itemize}

To get a better understanding of the extreme nature of this new class of
objects, we need more observations. In particular, estimations of the
albedo and of the surface fraction occupied by the lava ocean could
constrain the composition and the structure of the crust of the planet.
The challenging point lies in the difficulty of the direct observation
of the emitted/reflected light from these objects because of their very
small radius and of the extremely high contrast with respect to the star
they orbit. However, \citet{Rouan_2011} and \citet{Samuel_2011} have
shown, respectively, on the example of KEPLER-10b and CoRoT-7b, that the
modulation of the planetary emitted flux due to the phase effect should
be detectable in a near future with the James Web Space Telescope
(JWST). With that in mind, we decided to explore the variety of phase
curves that an observer based in the Earth neighborhood would see during
the orbit of the considered SE in different physical situations. We
decided to focus on the case of CoRoT-7b as a typical example of a very
hot SE around a sun-like star. One could easily extend this study with
the same method to any other very hot SE, while adapting the properties
of the system as the star and the planetary radia, their masses, the
magnitude of the star, the orbital period, and so on.

\section{Goal}

In this article, our goal is thus to estimate how precisely future
observations could help to constrain the physical properties of the
surface and atmosphere of very hot SE. More specifically, we question
the possibility of exploiting the reflection/thermal emission phase
curves of very hot SE if they are taken with the accuracy that is
anticipated for the JWST. Indeed, this space observatory will be
equipped with the largest telescope ever launched in space (6.5 meters
diameter) and will have, among its various instruments, a near-infrared
(IR) spectrometer (NIRSpec) with a $0.6-5\mathrm{\mu m}$ spectral range
that fits very well with this kind of observations.

In particular, we adress in \S\ref{part1}, \ref{part2}, and \ref{temp} the three following problems: \begin{itemize} \item
How precisely can we infer the surface albedo of the planet?  Can we
distinguish a synchronously rotating planet (as expected) from a more
rapidly rotating one? Can we estimate the rotation period of the planet?

\item Can we infer the presence of an atmosphere, and what is the
minimum detectable surface pressure?

\item Can we infer the planetary temperature map by inverting a low
resolution spectrum obtain with NIRSpec on JWST?

\end{itemize}

The range of parameters visited for this study is summarized in the
Table~\ref{table:3}. To avoid having too many free parameters,
we consider CoRoT-7b as a prototype hot SE and carry on the analysis in
this study case only. Furthermore, we focus on broadband photometry, so
the composition of the rock \citep{Hu_2012} or the observation of
molecules in a cometary tail, which requires high spectral resolution,
is not examined.

\section{Physics and models}

In this section, we describe the various physical processes that we have
taken into account in the physical model of the planet. Because it is
based on simple and conservative assumptions, we consider the 
\citealt{Leger_2011} model of an atmosphere-free, tidally-locked planet
as being the reference model, and we start by describing its properties in more
detail. Then, we discuss the consequences of relaxing
some of these constraints.

\subsection{The fiducial model: phase-locked, low albedo,
atmosphere-free, and lava ocean model}

As described in L11, when the planet is tidally locked and has no
atmosphere, the surface temperature is in thermal equilibrium and is
given at each position (defined by its zenithal angle $\theta$) by
\begin{equation} T_{\mathrm{surf}}=\Big( \frac{\epsilon_5}{\epsilon_2}
\Big)^{\frac{1}{4}} \Big( \frac{R_\star}{a} \Big)^{\frac{1}{2}}
\cos^{\frac{1}{4}}  (\theta) T_\star\,, \end{equation} where $R_\star$
and $a$ are  the radius of the SE and the star-planet distance,
respectively. The $\epsilon_5$ and $\epsilon_2$ are the emissivities of
the planetary surface in the wavelength range of a black body emission
at (resp.) 5250 K and  2200 K. If we assume these two values to be very
close ($\epsilon_2=\epsilon_5\equiv\epsilon$; L11), which is equivalent
to assuming that the albedo, $A=1-\epsilon$, is constant over the whole
visible/IR wavelength range, $T_{\mathrm{surf}}$ is independent of the
albedo of the planet ground.The study of L11 showed that such a situation would lead
to a large lava ocean, which is mainly composed of alumina: the most volatile
chemicals species must have evaporated and re-condensed over the
ocean shores. Assuming an alumina ocean composition, we can derive the
extension of the liquid-lava ocean with the shore that corresponds to the
melting temperature isotherm.

Now, if we assign any value to
$\epsilon_2=\epsilon_5\equiv\epsilon\equiv1-A$, we can derive the planet
to star flux ratio at any wavelength and any time along CoRoT-7b orbit.
Based on the measurement of Earth lava, L11 assumes a low albedo on
CoRoT-7b, but it is a surprise that Kepler-10b seems to show an
unexpectedly high Bond albedo of about 0.5 \citep{Batalha_2011,
Rouan_2011}. The visible light from the host star is more efficiently
reflected, but on the other hand, the thermal emission must be less
intense, as the albedo is the complementary of the emissivity.  If we
measure these two effects by observing the phase curve at different
wavelengths, we must be able to estimate the albedo value.

\subsection{Non synchronous rotation} 

Although tidal locking is the equilibrium rotation that is reached by a
fluid object subjected to tidal dissipation on a circular orbit
\citep{Hut80}, many physical processes can lead to a non synchronous
rotation, even when tidal friction is strong. For instance, if a third
body in the system (i.e. CoRoT-7c for example) is close and/or massive
enough to perturb CoRoT-7b orbit, the planet may not be in a perfectly
circularized orbit, which could result in a pseudo-synchronous (faster)
planetary rotation. Furthermore, if the solid planet exhibits a
permanent asymmetry, it could also be trapped in a spin orbit resonance
like Mercury, especially if some eccentricity is present \citep{MBE12}.
We only consider a coplanar spin-orbit in all this situations.

When this non synchronous rotation is considered, we use the duration of the day as a parameter to define it. This is the exact analog to the
solar day that lasts 24 hours on Earth. In the following, we use
the "day" to quantify the rotation of the planet, because it is the
important time scale for an irradiated point at the surface of the
planet. When we use the expression rotation period, we refer
to the day duration.

To model the temperature map, we need to take into account the thermal
inertia of the ground. For the sake of simplicity, this is done by assuming
that diffusion is efficient only within an inertia layer of limited
depth where temperature is homogeneous. In principle, the thickness of
this inertia layer depends on the rotation speed of the planet. As in any diffusion phenomenon, one can
assume that the depth of the heat
penetration must be proportional to the square root of the heating
duration. As the day is shorter (insolated time), then the thinner
the rocky inertia layer will be. If we consider of a rocky material with a 
diffusivity  $D \sim 1.2\,10^{-6}\mathrm{m^2 s^{-1}}$) and heating
with the periodicity P,  we can estimate the heat penetration depth by
\begin{equation} \Delta x = \sqrt{\frac{DP}{\pi}}\,. \end{equation}

However, to not mix the effect of the depth of the inertia
layer with the effect of the rotation period, we prefer to estimate
directly the inertia layer size. For a typical day of 20.5 hours (the
orbital period of CoRoT-7b), we derive a depth of about 15 cm. Thus, we
consider the three following heat penetration depths: 2.5 cm, 25 cm, and
2.5m. This covers the expected range for this parameter, while covering
the case of materials with much higher/lower diffusivities. Once this
depth is chosen, the thermal inertia of the layer is parametrized by the
specific heat of rock (which we assume to depend linearly on the
logarithm of the temperature from $\sim 50$ to $\sim 2.\,10^3 \mathrm{J
Kg^{-1}K^{-1}}$ from $70-2500 \mathrm{K}$; \citealt{Adk83}). A more
precise estimation of the thermal structure of the floor would
necessitate a particular study which is out of the scope of this work.

When the temperature reaches the liquefaction/solidification threshold,
it is held fixed until the whole material in the inertia layer is
molten/solidified. The latent heat used for this transition is $4 \times
 10^5 \mathrm{J \,kg^{-1}}$. The  duration  to melt/solidify the inertia
layer depends on the latent heat,  the intensity of the incoming stellar
flux, and the thickness of the rocky layer.

%, in proportion of  the power 4 of the temperature, according to the
%Stephan's law . 
%
   \begin{figure} \centering \includegraphics[width=9cm]{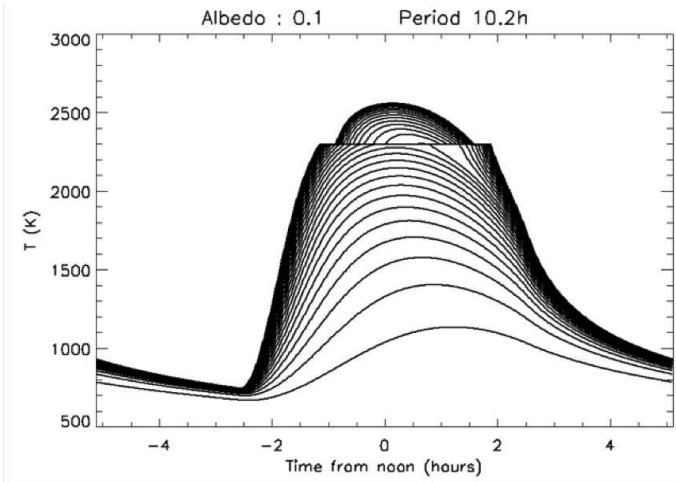}
   \caption{Temperature variation at different latitudes of a non
   phase-locked CoRoT-7b, with a hypothetical day of 10.2 hours (half
   the orbital period of the planet). As the planet is rotating, we can use the time or the longitude on the abscise axis.
   Here, the zero on the time axis corresponds to noon. Each curve
   represents the temperature at a given latitude (from $0$ to $85 $,
   with a step of $5 \ensuremath{^\circ}$).  The plateau at
   $T_\mathrm{fusion}=2300 \mathrm{K}$ is due to the extra energy needed
   to overpass the latent heat at the threshold of the melting
   temperature. We can interpret this graphic like this for a point
   which is at low latitude: the rock melts around 1.5 hours before noon
   and so the lava ocean appears; it fully goes back to solid state 
   more than 2 hours after noon. The  extension of the lava ocean is   $
   \pm 60\ensuremath{^\circ}$ in latitude} \label{fig1} \end{figure}

\subsection{Atmospheric redistribution}

A dense atmosphere can dramatically change the temperature map of the
planet. For example, the temperature difference between the day side and
the night side is almost 10 K on Venus and on the Earth, it is 300K
on the Moon or 600K on Mercury! In the case of very hot SE, the
presence of a dense global atmosphere would lead to the redistribution
of the heat by advection motion of the gas: the temperature contrast of
a few thousands of Kelvin which is expected in the Leger et al. model, could be
strongly reduced. Of course, such a scenario would requires us to explain how
an atmosphere can survive the combination of  a strong wind and of an
extremely intense UV field, but this cannot be totally excluded.

Our simulations of a planet with atmosphere have been performed using
the LMD generic global climate Model (GCM) specifically developed for
the study of extrasolar planets \citep{WFS10,WFS11,LFC13} and
paleoclimates \citep{WFM12, FWM12}. The model uses the 3D dynamical core
of the LMDZ 3 GCM \citep{HMB06}, which is based on a finite-difference
formulation of the classical primitive equations of meteorology. A
spatial resolution of 64x64x20 in longitude, latitude, and altitude was
used for the simulations.

Because the composition and mass of the putative atmosphere remains mostly unconstrained to date, we use a very simple approach and
consider only a gray gas. This simple approach is further supported because spectroscopy of complex molecules is poorly known
for the high temperatures reached on CoRoT-7b. Considering that our
purpose is to see whether or not the atmosphere's impact on the phase
curve would betray its presence, this case is conservative, since,
as shown by \cite{SWF11}, any molecular spectral feature in the spectrum
would more easily reveal the atmosphere.

Once a unique specific opacity ($\kappa$ in $m^2/kg$) has been
chosen\footnote{There is no clear distinction between the infrared and
optical part the spectrum in our case. Given the constraints available,
and  notwithstanding the inherent difficulty in choosing the cutoff
wavelength,  separating our opacities into spectral channels seems
uncalled for.} and we have defined the bands in which the flux is
computed\footnote{As fluxes are integrated over the whole stellar and
planetary spectrum in our gray case, the choice of the bands is
inconsequential for the state of the atmosphere.}, the model uses the
same two-stream algorithm developed by \cite{TMA89} to solve the plane
parallel Schwarzschild equation of radiative transfer in previous
studies. Thanks to the linearity of the radiative transfer equation, the
contribution of the thermal emission and downwelling  stellar radiation
can be treated separately, even in the same spectral channel. Thermal
emission is treated with the hemispheric mean approximation and
absorption of the downwelling stellar radiation is treated with
collimated beam approximation.

With respect to the other physical parameters, we consider an highly
idealized case. We do not account for any condensible species in the
atmosphere and, accordingly, do not consider any radiatively active
aerosols. The surface is considered flat with a constant albedo of
$A=1-\epsilon=0.3$. Turbulent coupling with the surface  in the
planetary boundary layer is handled by using the parametrization of
\cite{Mellor_1982} and convective adjustement is performed whenever the
atmosphere is found to be buoyantly unstable.

Given these assumption, and that planetary parameters are
known, we are left with only two free parameters: namely, the specific
gas opacity ($\kappa$) and the atmospheric column mass ($m$) or,
equivalently, the atmosphere normal optical depth ($\tau$) and the mean
surface pressure ($p_s$). Basically, these quantities are linked through
\begin{equation} p_s = m g\,, \end{equation} and \begin{equation} \tau =
\kappa  m = \kappa  p_s / g\,, \end{equation} where $g$ is the gravity.
Once these parameters are chosen, the model is run from a uniform state
at rest until an equilibrium is reached in a statistical sense. Emission
maps produced by the GCM are then processed by the tool described in the
previous sections to simulate the phase curve and quantify the
difference with atmosphere-free planet case.

\begin{table*} \centering                          % used for centering
table \begin{tabular}{c c c c c}        % centered columns (4 columns)
\hline\hline                 % inserts double horizontal lines Model &
Day duration  & Ground albedo & Atmospheric pressure &  Specific opacity
\\ \hline Phase locked, atmosphere-free & $\infty$  & from $0.1$ to
$0.7$ & $-$  & $-$\\    % table heading Rotating planet, atmosphere-free
 & from $10$h to $\infty$ & from $0.1$ to $0.7$ &  $-$  & $-$\\
Phase-locked, with atmopshere & $\infty$ & $0.3$ &  $10$ mb,  $100$ mb
and $1$ b & $1.5 \times 10^{-5} \leq \kappa \leq 1.5 \times 10^{-3}$\\
\hline                        % inserts single horizontal line
\end{tabular} \caption{Table of the different parameters used for each
model.} \label{table:3} \end{table*}

\section{From the model to the simulation of observations} 
\label{simofobs}
\subsection{Phase curve model and simulation of observations with JWST}

From the temperature maps computed from the aforementioned model, we can
derive the expected phase curve that an observer would measure from the
vicinity of the Earth by integrating outgoing flux over the visible
hemisphere. The expected planetary flux is the sum of the blackbody
emission and the reflected light.

We consider several wavelengths, chosen to optimize the signal to noise
ratio (SNR) of a supposed observation with the spectrograph NIRSpec on
the JWST. The choice of this instrument allows us to investigate a large
spectral range ($0.6$ to $5 \mathrm{\mu m}$) from the red, which is useful for
characterizing the reflected light to the mid IR, where the thermal
emission dominates. The spectral range is divided into five channels which is given
in Table~\ref{table:4} (obtained by binning properly the spectrum) and
allows us to study the phase curves as a function of the wavelength
(Fig.~\ref{fig3}).

\begin{table*} 
\centering                          % used for centering
\begin{tabular}{l c c c c c}        % centered columns (4 columns)
\hline\hline                 % inserts double horizontal lines 
Bands number & 1  & 2 & 3 & 4  & 5\\ \hline Bands limits ($nm$): min /
\textbf{center} / max & 600/\textbf{800}/1000  & 1000/\textbf{1125}/1250
& 1250/\textbf{1375}/1500 & 1500/\textbf{1750}/2000 &
2000/\textbf{3500}/5000 \\    % table heading

\hline                        % inserts single horizontal line
\end{tabular} 
\caption{Photometric bands used in the simulations of section \ref{simofobs}.} \label{table:4} \end{table*}

Then, we use the exposure time calculator\footnote{ETC is a product of
the Space Telescope Science Institute, which is operated by AURA for
NASA.} \citep{Sosey_2012} online tool for estimating the uncertainties on
the modeled measurement (instrumental and photon noise); we chose an
exposure duration of 72s, which gives us a good timing  for both the phase
variations at the different wavelengths, and for the detection of the
primary and secondary transits. The transits signature is modeled using
the depth from by L\'eger et al. (2009). As we want to exploit the most
precise phase-curve variations, we need at least one whole orbital
period (20h30min for CoRoT-7b) of observation. We simulate 41 hours of
continuous observation of the CoRoT-7 system (2 orbital period of the
SE). In such conditions, the peak-to-peak amplitude of the phase curve
is the most significant information for the observer and makes the estimation of the global albedo possible (\ref{alb}). Figure ~\ref{fig5} shows
typical light-curves obtained by this method. A precise analysis of
these phase-curve temporal variations is a more delicate task but can
provide valuable information on the rotation of the planet and/or the
surface temperature map (\ref{rot} and \ref{temp}).

Note that the secondary transit depth match directly to the peak-to-peak
amplitude of the phase curve variations because both the reflected light and the planet thermal emission are
masked when the star hides the planet, except if the thermal emission of the night side of the planet
is not negligible, in which the secondary transit can appear
deeper than the phase curve amplitude value.

\subsection{Inverse problem}

To quantify the information contained in the light-curves, we used a simple method to explore the simulated data: We fit a modeled
light-curve to the data and try to minimize a $\chi^2$ function, while
browsing the whole parameter space considered (for example, the
albedo-day duration space, when we explore the possibility of a non
phase-locked situation). We then compare the best couple of parameters
obtained with this method with the ones used to produce the simulated
data set. We derive the uncertainties directly from the $\chi^2$ map
(e.g. Fig.\ref{fig6}).

Finally, we try to answer the question: is this observed light-curve
compatible with the phase-locked airless planet model? To that purpose,
we compute the $\chi^2$ difference between the two curves as an
estimator of the likelihood. We can then give a simple answer: the simulated situation is either distinguishable or not from the L11's
one.

The global results are presented in the following, as well as few 
specific exemples to illustrate our discussion.

  \begin{figure} \centering \includegraphics[width=9cm]{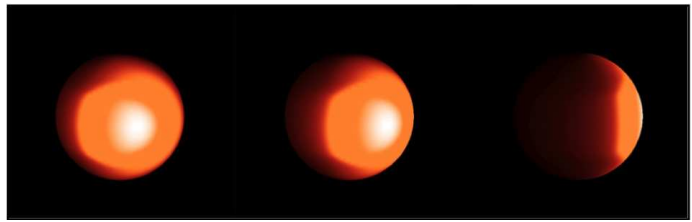}
  \caption{Those three pictures show an example of the thermal emission
  map of a non phase-locked SE at three consecutive value of the
  orbital phase. The asymmetry in longitude translates the thermal
  inertia of the ground, partly due to the latent heat when the melting
  of the rock occurs. } \label{fig2} \end{figure}

  \begin{figure} \centering \includegraphics[width=9cm]{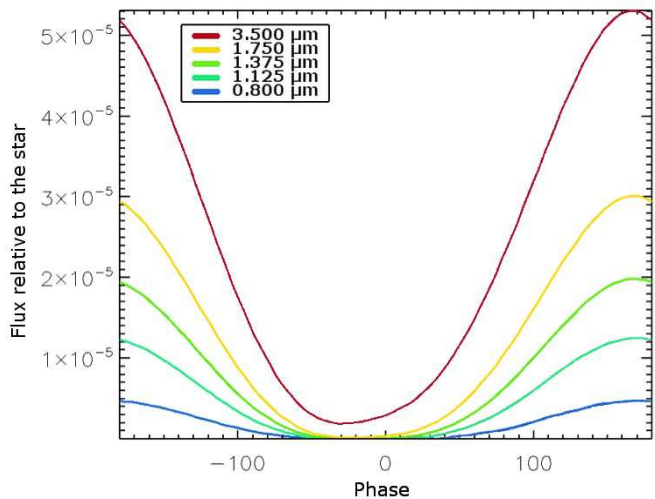}
  \caption{Phase curves in different spectral bands for a non
  phase-locked CoRoT-7b with a hypothetical day of 10.2 hours and an
  albedo  of 0.1. From purple to red, a wavelengths of resp. $0.8
  \mathrm{\mu m}$, $1.125 \mathrm{\mu m}$, $1.375 \mathrm{\mu m}$, $1.75
  \mathrm{\mu m}$, and $3.5 \mathrm{\mu m}$.  The amplitude has been
  normalized by the stellar flux. As expected (II.4), the  phase curve
  at a longer wavelength shows a clear delay with respect to the one at
  the shortest wavelength. The zero on the phase axis  corresponds to
  the secondary transit (the observer-star-planet are aligned in this
  order). The transit and secondary transit signals are not modeled on
  this figure.} \label{fig3} \end{figure}

  \begin{figure} \centering \includegraphics[width=9cm]{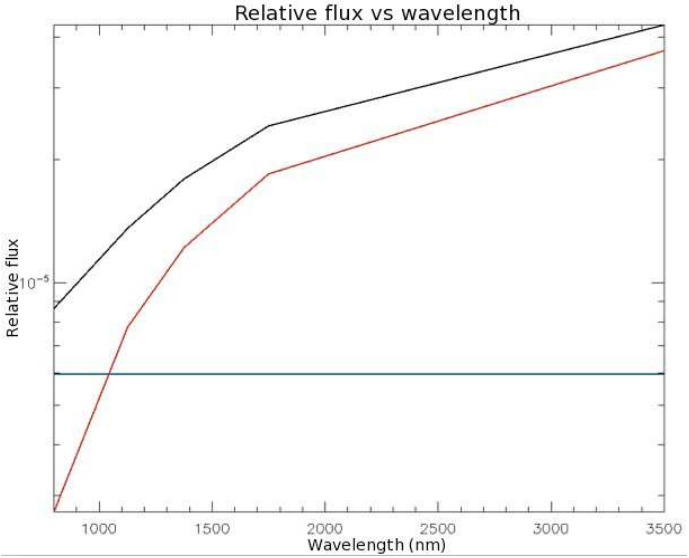}
  \caption{Contribution to the phase curve variation of the reflected
  light (blue line), thermal emission (red) and total (black). In this
  example, we took an albedo value of 0.5, so that the reflected flux
  from the star is higher than the emitted thermal flux from the planet
  only at wavelengths lower than 1000nm.} \label{fig4} \end{figure}

  \begin{figure*} \centering \includegraphics{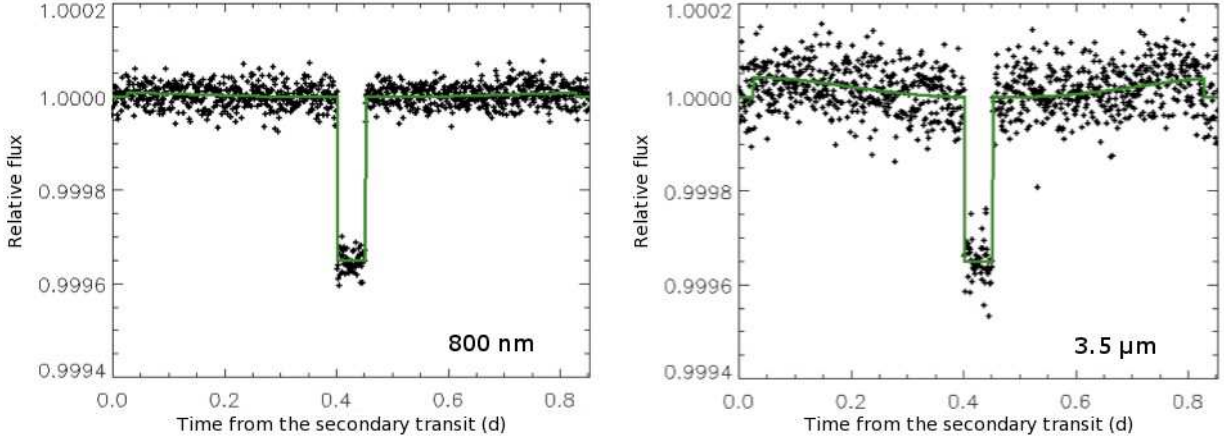}
  \caption{Simulated light curves corresponding to two extreme 
  wavelengths of the observation using JWST-Nirspec. On the left, the
  ideal light-curve (green solid line) expected at 800nm and the same
  with the added noise (black cross). The same symbol and color code are
  used on the right figure for a wavelength of $3.5\mathrm{\mu m}$.}
  \label{fig5} \end{figure*}

\section{Results }

\subsection{Retrieval of albedo and rotation} \label{part1}

In this subsection, we consider only the airless case and focus on the
quantification of the constraints put by light curves on the [albedo,
rotation period] couple. To that purpose, we simulated noisy light
curves for different values of the albedo and rotation period. For each
given noisy light curve, we then mapped the value of $\chi^2$ for the
whole [albedo, rotation period] parameter space. Examples of such
$\chi^2$ maps are shown in Figures~\ref{fig6} and \ref{fig7}. Contours
in these maps show us what part of the parameter space can be ruled out
for any given confidence level.

 We extended this study to different situations with a variable heat
 penetration depth at the surface: the deeper the heat penetrates the
 planet crust, the more important the thermal inertia will be. This
 moderates the temperature variations and lengthen the associated delay.
 As a result, we see that larger heat penetration depth entails
 tighter constraints on the rotation period.

\subsubsection{Albedo} \label{alb}

\begin{figure} \centering \includegraphics[width=9cm]{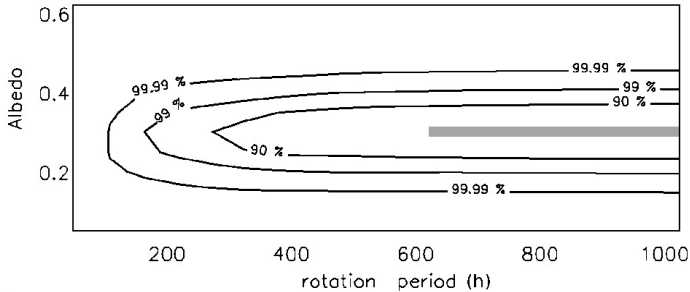}
  \caption{$\chi^2$ map showing the compatibility between the
  observation of a phase-locked CoRoT-7b with a 0.3 Bond albedo and a
  family of models with various possible albedos and rotation periods, with a inertia-layer thickness of 25cm. The different contours
  correspond to the regions where the confidence level for the
  phase-locked model is 90\%, 99\%, and 99.99\% in the two-parameter
  space. The gray rectangle corresponds to usual $1 \sigma$
  incertainties (68\% confidence level) when we independently study each
  one of the parameters, while the other is held fixed.} \label{fig6}
  \end{figure}

To focus on the retrieval of the albedo, we first consider only fiducial
models with a synchronously locked rotation. An example of a $\chi^2$ map
for the case with an albedo of 0.3 is shown in Figure~\ref{fig6}. From
this map, we can see that the retrieval of the albedo and rotation
period is not correlated to each other. We can thus infer a $1\sigma$
error bar on the retrieved albedo.

The detailed results are given in  Table ~\ref{table:1}. The  albedo
corresponding to the maximum confidence level  are given for the
different simulations. The simulated Bond albedo is allowed to vary from
0.1 to 0.7. We remind that we assume that the albedo is
constant with a wavelength in the whole range. A finer hypothesis is discussed in \S\ref{temp}. The accuracy on the retrieved  albedo is
better than 0.05, which is a satisfactory result if we consider that the
estimated albedo of the rocky planets is a key factor to get
information on the nature of the most superficial layer at the surface.
For instance, we should be able to confirm with a much better accuracy the surprising high albedo
derived for Kepler-10b \citep{Rouan_2011, Batalha_2011}.

\begin{table*}
\centering                          % used for centering table
\begin{tabular}{c c c c c}        % centered columns (4 columns)
\hline\hline                 % inserts double horizontal lines
Inertia layer thickness (cm)  & \multicolumn{4}{c}{Simulated albedo}\\
 & $0.1$ & $0.3$ & $0.5$ & $0.7$\\    % table heading 
\hline                        % inserts single horizontal line
   2.5  &$a= 0.05 \pm 0.03$  & $a=0.30 \pm 0.03$  & $a= 0.50 \pm 0.03$ & $a=0.70^{+0.07}_{-0.03} $  \\       
         &$P_{min}=27$h&$P_{min}=27$h&$P_{min}=27$h&$P_{min}=27$h  \\
 & & & & \\
   25  &$a= 0.15 \pm 0.03$  & $a= 0.3 \pm 0.03$  & $a= 0.55 \pm 0.03$ & $a= 0.65 \pm 0.03$  \\       
        &$P_{min}=81$h&$P_{min}=297$h&$P_{min}=27$h&$P_{min}=27$h  \\
 & & & & \\
   250 &$a= 0.10 \pm 0.03$  & $a= 0.30 \pm 0.03$  & $a= 0.50 \pm 0.03$ & $a= 0.70 \pm 0.05$ \\       
        &$P_{min}=836$h&$P_{min}=620$h&$P_{min}=556$h&$P_{min}=674$h  \\

\hline                                   %inserts single line
\end{tabular} 
\caption{Simulated observation with NIRSpec of a
phase-locked CoRoT-7b, followed by a phase curve fitting. Different
albedos and different depths of heat penetration are used. For each case, the table gives  $a$, the Bond albedo  of the planetary surface, and $P_{min}$, the shortest rotation period of the planet compatible with the simulated data (with a 68\% probability, equivalent to the usual $1 \sigma$ uncertainy).} \label{table:1} \end{table*}

%%%%%%%%%%%%% mardi *%%%%%%%%%%
%%%%%%%%%%%%% 
\subsubsection{The planet rotation} \label{rot}

When the rotation is not synchronized, the planetary surface receives a
variable stellar intensity during the day, and its temperature evolves
with time. As on Earth, the maximum temperature does not occur exactly
when the star is at its zenith (at noon) but at a short time later:
because of its thermal inertia, the ground temperature still increases
until the absorbed incoming stellar flux becomes lower than the power
lost by thermal emission. Furthermore, if the rock temperature exceeds
$T_\mathrm{fusion}$, then the ground will be momentarily melted, and the
latent heat will be an additional parameter that changes the thermal
inertia. In these conditions, we can get an ephemeral lava ocean that 
appear and disappear periodically at the same hour everyday, provided
that cooling is fast enough, as illustrated on Fig. \ref{fig1}.

Finally, one can notice that a given point of a
non phase-locked planet sees its temperature increase as the stellar
light intensity increases during the "morning". Then it cools down
during the "afternoon" by emitting radiations ($\sim \sigma T^4$ ), but
the temperature decreases more and more slowly (as $T^4$ decreases
very quickly), making an asymmetric temperature evolution with respect to noon.

A non synchronous planet with a finite thermal inertia thus shows a
signature of its rotation in its thermal emission phase curve. The
maximum of the IR phase curve does not occur at the same time than for
the reflected light. This produces a peculiar signature that would
indicate a non synchronized rotation of the planet (see
Fig.~\ref{fig2},~\ref{fig3}). Since the contrast between the
phase-curves (at different wavelength) depends on the albedo, the
accuracy of the rotation determination depends on this parameter.

As a result, there is always a rotation period below which the thermal
phase curve is too distorted to be confounded with the one of the phase
locked planet (to a given confidence level). This lower rotation period
(at 1$\sigma$) is also given in Table~\ref{table:1}. In the case where
the thermal inertia is low (heat is exchanged only in the first 2.5cm of
the ground), one cannot differentiate a phase-locked planet from a rotating planet with a 20
hour period. On the other hand, the larger (and more
realistic) the inertia-layer thickness we use from 25 to 250 cm, the
larger this minimum rotation period is. As visible in Fig.~\ref{fig6},
(simulated albedo of 0.3 and a thermal exchange thickness of the surface
of 2.5 m), we see that we can discriminate a rotating planet with a 200
hour period  from a phase-locked one with a confidence level of 99\%.

It can also be noted that we can see a clear correlation between the admissible
range of rotation period and the albedo, even if the albedo value is always well
constrained: for the lowest Bond albedo
values, a larger range of  rotation period  is excluded. For a 250 cm
inertia-layer thickness, an albedo of 0.1 allows us to exclude  the
rotating planet scenario for a period lower than 840 hours (i.e. almost
42 orbital periods of CoRoT-7b). In the same conditions with
an albedo of 0.5, this limit is nearly 550h (27 orbital periods).
Finally, we notice a few outliers in these results that do not follow
these global trends. This is  probably due to  statistical deviations in
the random noise sample of our simulations, since each result
corresponds to one single simulation.

Going further, we tried to quantify the accuracy with which the rotation
period of a non synchronous planet could be retrieved from the light
curve. Table~\ref{table:3} shows the rotation period limits below
which we can significantly distinguish a rotating planet from a
phase-locked one. We can deduce that we would be able to estimate a most probable value for the day duration, if the observed planet was showing
a short rotation period, which is a day duration shorter than those limits.

\begin{figure} \centering \includegraphics[width=9cm]{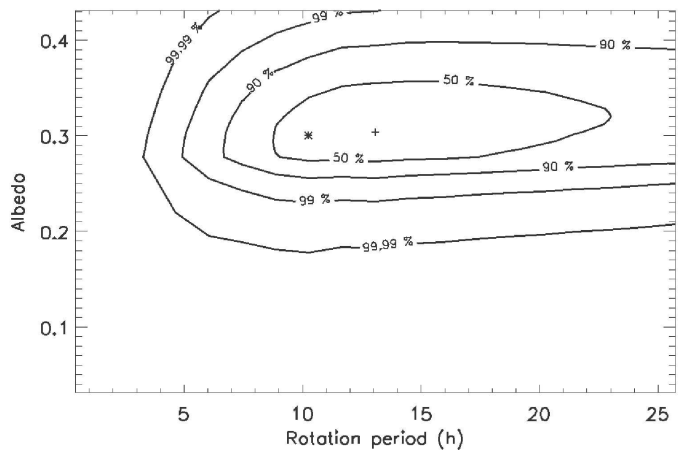}
  \caption{ $\chi^2$ map comparing a simulated  observation of a rotating  CoRoT-7b with a 0.3 Bond albedo to a family of models with
  various albedo, rotation periods, and a inertia-layer thickness of
  25cm. The cross is at the location of the best fitted model in the
  albedo-period space, and the star marks the position of the initial 
  model (albedo=0.3 and rotation period $=10.25$h $= 1/2\,$orbital
  period of CoRoT-7b).} \label{fig7} \end{figure}

Again, we clearly find that the thickness of the inertia layer strongly
influences the ability to retrieve the rotation period. We also notice
that the more the
phase-curve is singular as the planet rotates upon itself faster and the easier the determination of the period
is. In very favorable cases (low albedo, short rotation period and large
inertia layer thickness), we can precisely constrain the rotation
period and the albedo of the simulated planet. An example of such a case
is given on Fig.~\ref{fig7}. The albedo is 0.3, the rotation period
(day) is 10.25h long (1/2 orbital period) and the heat penetration depth
is 2.5m. The best fit gives a Bond albedo of $0.3^{+0.06}_{-0.02}$ and a
rotation period of  $13.0^{+7.7}_{-2.1}$h. In this particular case, it
would be easy to precisely deduce the rotation period of the planet and 
the albedo; the phase-locked scenario could be rejected with a
confidence level of $13 \sigma$.

 Note that we did not consider  a high global temperature of the planet,
 as the possible mechanism we propose to explain a non phase-locked
 planet should imply\footnote{Even a slight non synchronization of the
 spin-orbit could lead to a global strong warming of the planet and
 could possibly lead to a fully melted planet (Franck Selsis, private
 discussion). But in this case, once the rock is liquid, the energy
 dissipation due to the tidal forces should decrease strongly so that
 the most superficial layer (which radiates the energy flux to space) 
 could then come back to a solid state, forming a crust.   The thickness
 of the crust would be such that an equilibrium of power is reached
 between some tidal dissipation within the crust, the diffusion of the
 heat of the melted subsurface and the outward radiation. The important
 point is that the planet surface could then have a solid surface, and a low tidal dissipation power. We did not conduct a full physical
 evaluation of this situation, and of course one will have to check if
 the time constant to return to equilibrium (i.e. phase-locked rotation)
 is sufficiently long so that there is a chance to observe the
 transition phase.}.  If it was the case, this could allow us to observe
 even more easily a potential signature of this situation, with a
 secondary transit significantly deeper than the phase-curves
 peak-to-peak amplitude, due to a lower day-night thermal contrast. In
 this state of mind, \cite{Selsis_2013} have recently explored the
 effect of the tidal heating on the light curves in more tricky
 situations, such as excentric orbits, spin-orbit resonances, and planets
 in pseudo-synchronized rotation.

\subsection{Atmospheric signature} \label{part2}

We now try to assess the possibility of detecting the presence of an
atmosphere from the analysis of the phase curves. To avoid mixing different effects, we consider only the phase-locked case. While the
presence of a significant atmosphere of silicates in equilibrium with
the magma does not seem to be favored (L11), one cannot totally rule
out yet the possibility of a non condensable gas being released by the magma
ocean, or created by photochemical processes \citep{SF09,CM11}. For
example, the presence of gaseous N$_2$ is not thermodynamically
impossible on the day side but also on the night side since the triple
point is around 63K and 13 kPa, when we expect a temperature between 50K
and 75K in the coldest region (even without atmospheric redistribution).
If the flux of the non condensable species released by the different
processes can compensate the atmospheric escape, we would be in presence
of an atmosphere in an out of equilibrium but steady state situation. In
any event, it is worth trying to
quantify the constraints on the presence of a potential atmosphere that
could be put by observations of phase curves, from an observational point of view.

\subsubsection{Simulations}

In a first set of simulations, we ran the model for surface pressures of
10mbar, 100mbar, and 1bar and kept $\kappa$ constant to have an optical
depth of $10^{-3}$, $10^{-2}$, and $10^{-1}$ respectively. This
represents an optically thin case. As expected, the principal
effect of the atmosphere is to smooth the initial surface temperature
contrasts between the two hemispheres of the planet. At high altitude,
hot air is transported from the day to the night side where it can radiate some
of its thermal energy.

  \begin{figure} \centering \includegraphics[width=9cm]{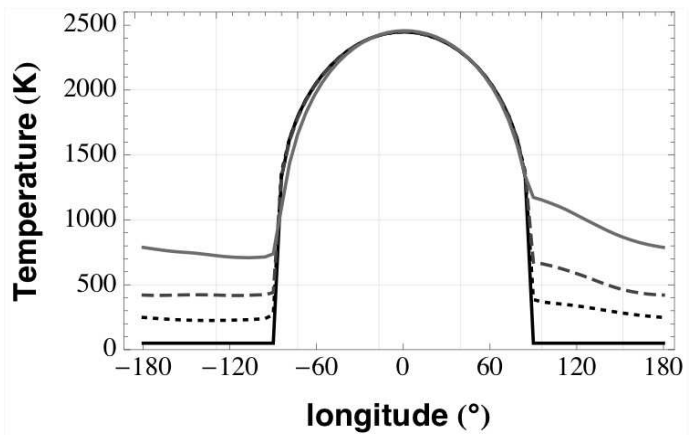}
  \caption{Temperature at the equator of CoRoT-7b in the atmosphere-free
  case (solid black) compared to the 10, 100, and 1000 mb gray atmosphere
  cases (dotted, dashed, and solid gray curves, respectively). Substellar
  point is at longitude $0$. The east/west asymmetry on the night side
  is due to the eastward winds at the equator.} \label{fig8}
  \end{figure}

\begin{figure} \centering \includegraphics[width=9cm]{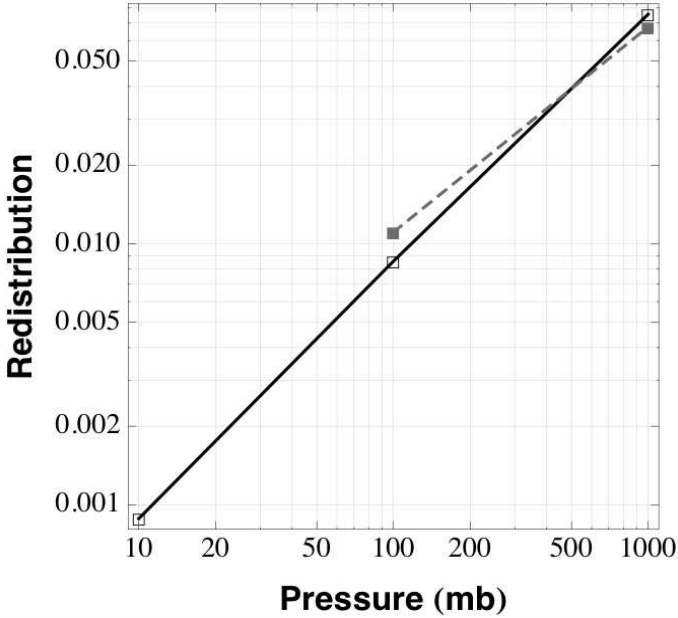}
  \caption{Bolometric redistribution efficiency factor as a function of
  pressure computed as the luminosity of the night side over the
  luminosity of the day side for the thin case ($\kappa=1.5 * 10^{-5}$;
  black solid curve). For comparison, redistribution in the thicker case
  is shown ($\kappa=1.5*10^{-2}$, yielding an optical depth of 10 for
  the 1bar case; gray dashed curve).} \label{fig9} \end{figure}

To estimate the impact of this redistribution on the light curve we
computed a {\it bolometric} redistribution efficiency ($\eta$) defined
as the ratio of the total luminosity (due to thermal emission) of the
night side over the luminosity of the day side. This ratio should be
near 0 for the atmosphere-free case and 1 if the emission is isotropic.
As can be seen on Fig.~\ref{fig9}, the redistribution efficiency has a
low value of 0.1\% for a 10mb surface pressure but rises up to
8\% for a 1b atmosphere in an almost linear fashion. This low
redistribution comes from the thermal timescale of the
atmosphere that scales as $T^{-3}$ and can be very short for such heavily
irradiated planet as CoRoT-7b. Simulations of planets receiving a less
intense insolation are indeed found to have a higher redistribution
efficiency (Leconte et al in prep).

We tested the impact of the optical depth at fixed surface pressure by
changing the opacity. Our results suggest that changing the opacity is
only  a second order effect when compared to the pressure dependence of
the redistribution. This is mainly because we use a
completely gray model. Emission arises at the same level as absorption
(typically at $\tau = 1$) which is the main driver of the day-night
temperature gradients. To first order, this should also be true in the
real case because stellar insolation and thermal emission occur nearly in the same
spectral regions at the high temperatures reached on the dayside. Therefore upper atmospheric levels are very hot; they have a
short radiative timescale and cool efficiently before reaching the night
side through transport mechanisms (like winds). As a consequence, the
optical depth does not directly affect the heat redistribution but
changes the amplitude of the phase curve: the thermal contrast increase
with the opacity of the atmosphere.

\begin{figure} \centering \includegraphics[width=9cm]{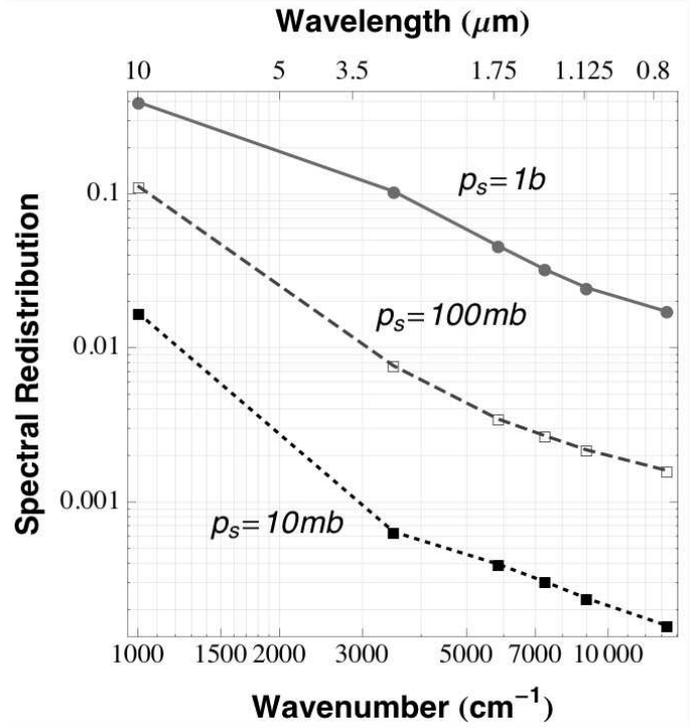}
  \caption{Spectral thermal redistribution efficiency factor as a
  function of wavenumber or wavelength,which is computed as the luminosity of
  the night side over the luminosity of the day side in each band for
  the thin case ($\kappa=1.5 * 10^{-5}$). The three different pressures
  are shown (10mb: dotted; 100mb: dashed; 1b: solid). The last band on
  the left (far IR) is not observable with JWST.} \label{fig10}
  \end{figure}

Finally, before assessing the detectability of an atmosphere directly 
from our synthetic light curves, we computed a spectral redistribution
as the luminosity of the night side over the luminosity of the day side
in each band of the model (where we have added one band to cover the
reddest part of the spectrum where low temperatures night side regions
emit strongly). This gives us an understanding of which channels carries the strongest atmospheric signature. Figure~\ref{fig10} shows us
that this redistribution is much more difficult to observe
 for wavelength shorter than 5 microns (less than 10\% for the 1b case) while the atmosphere can efficiently smooth luminosity contrasts in
the far infrared. 

\begin{figure*} \centering \includegraphics[scale=0.5]{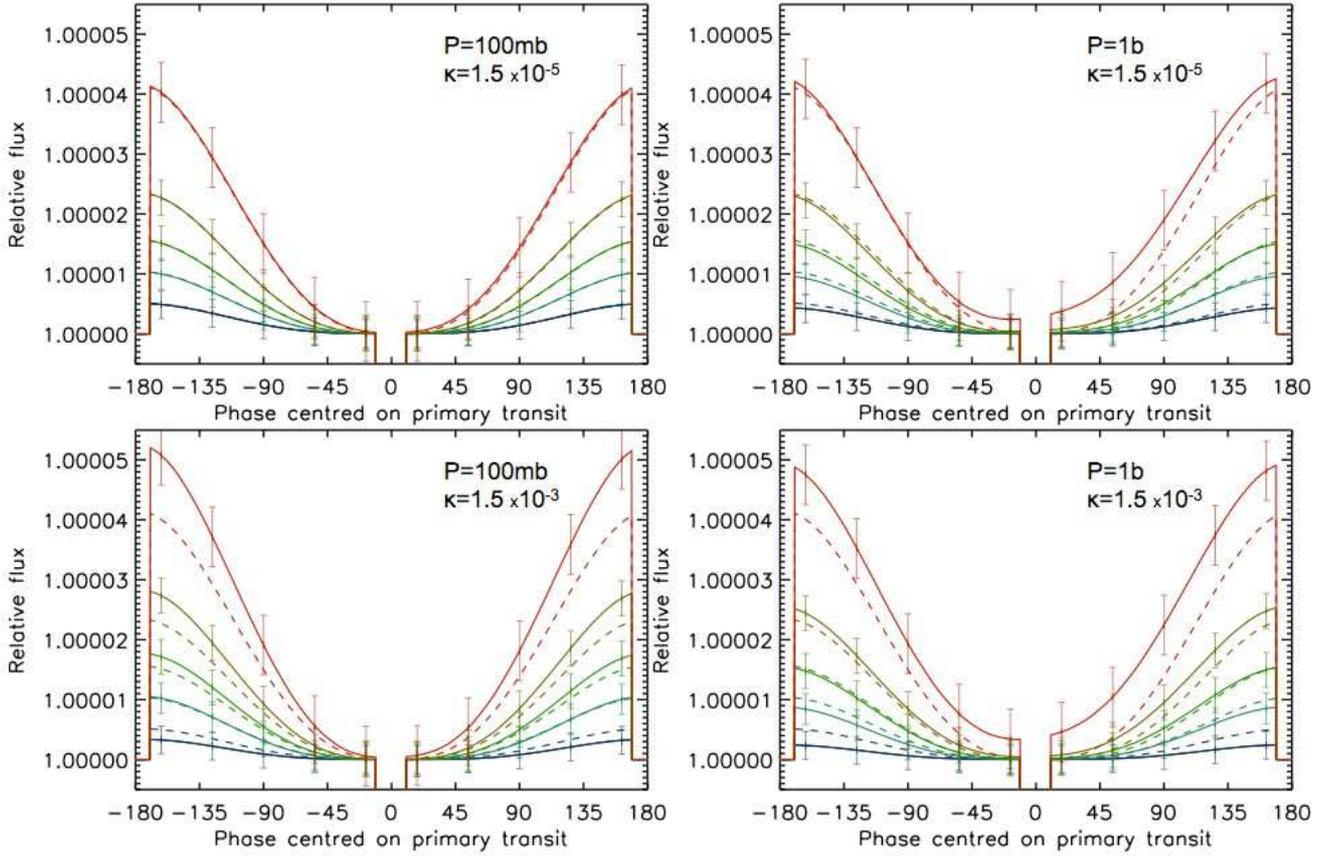}
  \caption{In these graphics, the five dashed lines represent the phase
  curve expected for a phase-locked atmosphere-free CoRoT-7b with an
  albedo of 0.3 in the five channels from $0.8 \mathrm{\mu m}$ (blue) to
  $3.5 \mathrm{\mu m}$ (red). The solid lines represent different
  situations with a grey atmosphere. The relative flux is plotted as a
  function of the phase, and the zero phase reference is defined as the
  primary transit position. The parameters for the atmosphere are: Top
  left:  P=100mb , $\kappa=1.5 10^{-5}$. Top right: P=1b , $\kappa=1.5
  10^{-5}$. Bottom left:  P=100mb , $\kappa=1.5 10^{-3}$. Bottom right:
   P=1b, $\kappa=1.5 10^{-3}$. The IR flux  from the cold regions is
  strongly correlated to the pressure, due to the heat transportation by
  the atmosphere. The opacity plays a role on the flux from the warmer
  part of the surface: as the atmosphere becomes more opaque, the substellar region becomes warmer; and the thermal flux is higher when these regions are in front of the observer (around phase 180).}
  \label{fig11} \end{figure*}

Taking into account the noise level expected with JWST, we have modeled
the effect of the atmosphere on the phase curves and compared it to the
atmosphere-free model. The difference becomes significant at a level
that crosses the $3 \sigma$ threshold just below 1 bar when we take
$\kappa=1.5\times  10^{-5}$. If we assume a higher opacity, we  find
situations more and more favorable from the detection point of view. As
shown on  Fig.~\ref{fig11}, downwelling stellar radiation is more efficiently absorbed and
thermalized before reaching the surface when the optical depth increases. As a result, less energy is
reflected in the blue part of the spectrum (hence the shallower phase
curve in this part of the spectrum) and more is re-emitted thermally in
the red part where the planet to star brightness ratio is higher. The
atmosphere lowers the bond albedo of the planet when the optical depth
is significant, and this amplifies the day-night thermal emission
contrast. On the other hand, the heat
transport efficiency grows and leads to the warming of the coldest part
of the planet surface for higher atmospheric pressures. When opacity and pressure are increasing, this
leads to two effects on the amplitude of the phase curves:  since both
the day and night thermal flux varies in the same direction, the
day-night contrast does not change strongly and  the phase curves are
shifted toward higher value.

In the high pressure and high opacity cases, one can note that, the
symmetry of the phase curves is broken, attesting to the effect of
(eastward) jets that transport heat, thus offering another signature of
the presence of an atmosphere. As a consequence, the observer sees a
higher flux from the planet in the interval of time between the 
transit and the secondary transit (positive phase) than during the
period that follows the secondary transit and precedes the primary one
(negative phase).

\subsubsection{Discussion}

In the most favorable situations, we conclude in a positive way on the
possibility to distinguish between the presence or the lack of an
atmosphere on CoRoT-7b (or on very hot super-Earths in equivalent
conditions):  if the pressure is about 1 bar or more and the specific
opacity is higher than  $10^{-5}$. The lava ocean model proposed by L11
does not predict such an atmosphere and the JWST should help to confirm
this assumption.

What would be the next step to improve the capability to detect the
signature of an atmosphere?

First, we did not include any spectral features in the gas
absorption spectrum because of the lack of constraints on the atmospheric
composition. This is consistent in that thermal
emission and stellar radiation spectral  domains are almost confounded
on the day side. However, the presence of opaque and transparent window
regions could directly be seen in the emission spectrum \citep{SWF11}.
In addition, we note that the stellar light could always heat the lower
altitude regions through transparent spectral windows which would allow the
upper part of the atmosphere to be cooler and thus to have a more
homogeneous horizontal temperature distribution, as explained before.
The emission would then be more homogenous in the opaque regions of the
spectrum than in the transparent ones \citep{SWF11}.

Second, the latter would not only carry sensible but also latent
heat if condensable species are present in the atmosphere (silicates
for example). As this process can be very efficient even when the amount of
condensible gas transported is very small, it could greatly enhance the
redistribution efficiency with respect to the one we have calculated.
However, considering the lack of constraints on the source and
composition of the atmosphere, this issue is beyond the scope of this
study.

Even if an atmosphere with a surface pressure between 10 and
100mb can reach the nightside temperature above the temperature of the
freezing point of water, note that these does not mean that such highly irradiated
planets are habitable because of the strong positive radiative feedback
of water vapor. To really assess the possibility of water stability at
the surface,  inclusion of the water cycle, as in \citet{LFC13}, is
mandatory.

\subsection{Surface temperature distribution.} \label{temp}

In the previous sections, we have shown that the future JWST observations will allow
us to retrieve global properties of the planet (in particular, albedo
and rotation rate) from the light curves. However, we had to assume a specific physical model to infer temperature maps from a
given set of these global parameters. In the following section, we relax
this assumption, and make an attempt to quantify how the temperature map
itself can be retrieved from the observations.

\subsubsection{Procedure}

To that purpose, we use the following hypotheses:

\begin{itemize}

 \item We assume that the photometric data are obtained using  the
 future JWST  NIRSPEC instrument with a proper smoothing in the
 spectral domain so as to synthesize a set of bandpass filters with a
 given signal to noise ratio. We used the exposure-time-estimator (ETC)
 available on the web for the evaluation of fluxes and noise. In the
 model, the minimum wavelength and the maximum wavelength are adjustable
 and the number of synthesized filters is adjusted so as to have
 $\Delta\lambda/\lambda = 0.2$. We look for the set of parameter values
 that allow us to retrieve the actual temperature distribution
 accurately enough at the lowest cost in terms of observing time. \item
 Any actual temperature distribution along a meridian (whose polar axis
 is the star direction) can be described with a fair approximation by
 $T(z)$, a polynomial expression of fourth degree, with respect to $z$, the
 zenith angle of the star direction. This angle is measured from the
 substellar point and increases toward the terminator. We indeed found
 that a polynomial expression of fourth degree can fit the radiative
 equilibrium temperature with an accuracy better than $10^{-3}$. If the
 temperature is redistributed through some other physical processes and
tends to be homogeneous, then the polynomial fit should of
 course still be valid. This assumption reduces the number of free
 parameters in the fitting procedure to five (the coefficients of the
 polynomial). We, however, also studied the case where a break in the
 temperature distribution arises because of the presence of the lava
 ocean and of a possible difference in the emissivity ratio
 $\epsilon_{\mathrm{IR}} / \epsilon_{\mathrm{vis}}$ between the lava
 material and the solid ground. \end{itemize}

The procedure we used is as follows. We first model the temperature
distribution of the planet assuming that it is phase-locked and that
the temperature at its surface is simply the one given by solving the
radiative equilibrium equation for a given albedo. Our model for the
irradiation takes into account the finite angular extension of the star.
The formulation is the one described in L\'eger et al. 2010. This is the
basic temperature distribution we wish to inverse.

Then we compute the overall spectrum of the planet as seen from a
distant observer. This provides the fluxes in the set of synthetized
filters. To assess the importance of the thermal infrared, we have
considered two cases, where the overall wavelength range is $0.7$ $-$ $3 \mu
m$ and $0.7$ $-$ $4.5 \mu m$.

The noise is added according to JWST prescriptions to obtain the desired
SNR. Finally, a Powell procedure is used to find the temperature curve, as
described by a fourth degree polynomial that produces the best fitting
spectrum.

\subsubsection{Results}

\begin{figure*} \centering \includegraphics{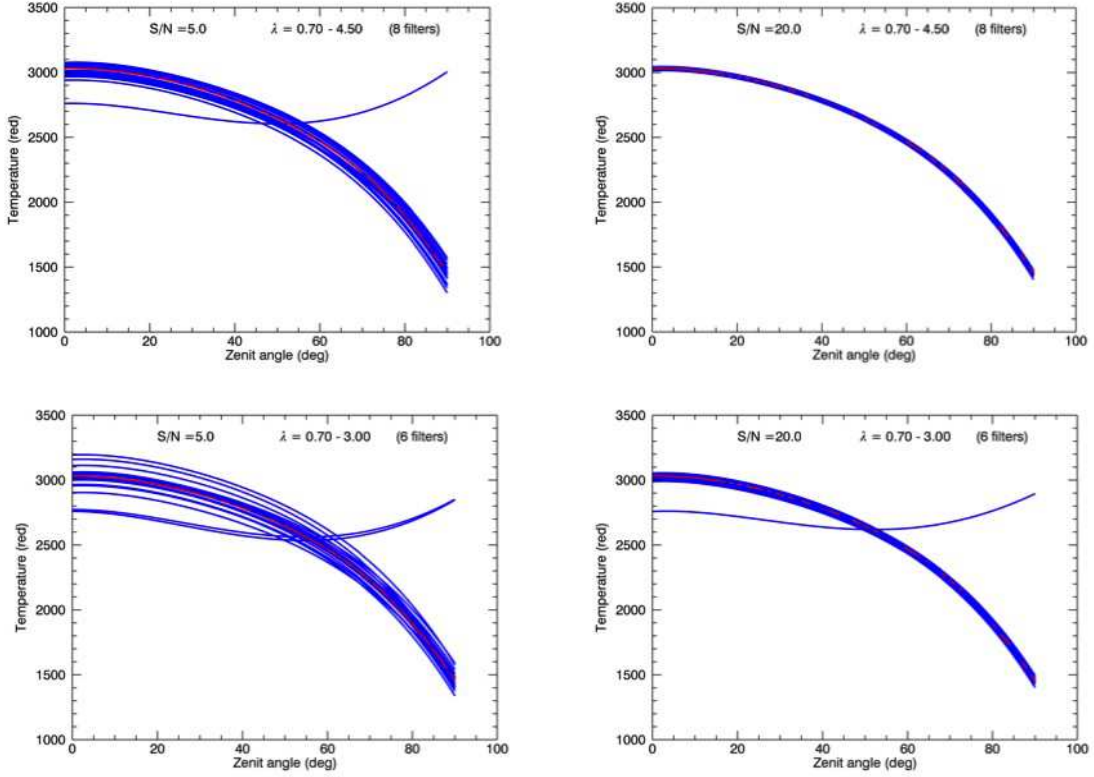}
  \caption{Four examples of temperature distribution retrieval from
  realistic observations of the planet spectrum using the JWST-NIRSPEC 
  instrument. For each case, 20 trials were done (blue curves). The
  temperature distribution to be retrieved corresponds to the red curve.
   The label indicates the conditions in terms of SNR (at the longest
  wavelength) and of wavelength range. In a few cases, one notes that one
  or two catastrophic solutions are found. For a SNR = 20 and a
  wavelength range extending to $4.5 \mathrm{\mu m}$, the accuracy on
  the retrieval of the initial distribution is good (< 10 K), and no
  catastrophic solutions is found. } \label{fig12} \end{figure*}

After a few trials, it appeared that the retrieved temperature
distribution was often fairly close from the initial one, except for a
few cases where the result was catastrophically different. In some
cases, for example, the temperature was increasing with solar zenith
angle instead of decreasing as it should (see  Fig.~\ref{fig12}). In
general, there is no intermediate situation: a fit is either fairly good
(say better than 10 percent everywhere) or frankly unacceptable
(worse than 500 K difference in a significant part of the curve).

We conclude that the requirement on the SNR is finally less based on the
accuracy of the retrieved temperature than on the probability to obtain
a catastrophic solution for which there is no way to validate it without excluding non physical solutions or by redoing the measurement
sequence. We show four examples of temperature
inversion for different sets of parameters in Figure~\ref{fig12}, where the SNR and the
wavelength range were changed. We give the
median error on the retrieved temperature\footnote{The computed 20 solutions of temperature distribution are deduced from simulated JWST-NIRSPEC observations. The added random noise is different for each.} and  the fraction of
catastrophic solution (on 20 trials) for different values of the
parameters in Table~\ref{table:2}. Even with a medium SNR of 7.5 and provided that the used
wavelength range extends up to $4.5 \mathrm{\mu m}$, the temperature
distribution at the surface of the planet could be retrieved with an accuracy of typically 30 K and a low risk  of unreasonable solutions.

\begin{table}      % is used to refer this table in the text
\centering                          % used for centering table
\begin{tabular}{c c c c c}        % centered columns (4 columns)
\hline\hline                 % inserts double horizontal lines SNR 
 & 5& 7.5& 10 & 20 \\ \hline 
$0.7-4.5\mathrm{\mu m}$ & 52.3 [.05] & 26.8 [0] & 20.8 [0]& 9.0 [0] \\ 
$0.7-3.0\mathrm{\mu m}$ & 64.4 [.10]&	30.6[.15]&	32.4[0]&14.4 [0.05]\\

\hline                                   %inserts single line
\end{tabular} \caption{Median error on the  temperature for different values of the SNR (from 5 to 20) and for two maximum wavelengths ($3$ and $4.5 \mathrm{\mu m}$).Within brackets: the fraction of catastrophic solutions.}\label{table:2} \end{table}

\begin{figure*} \centering \includegraphics{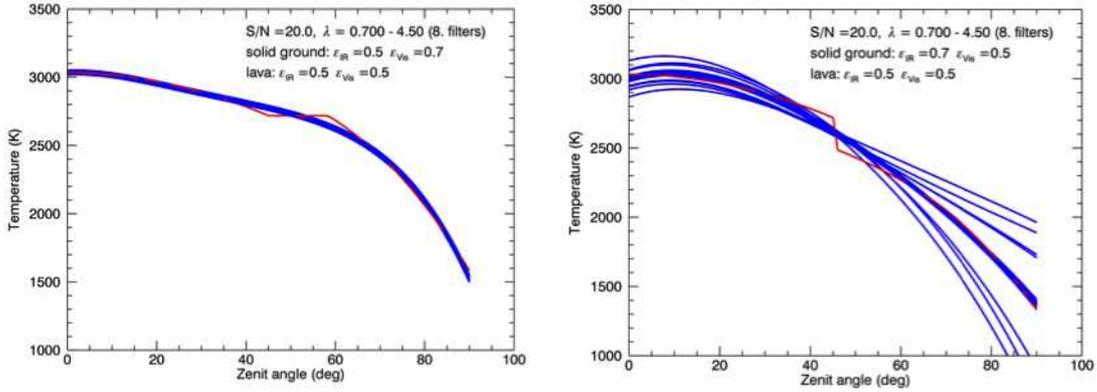}
  \caption{Examples of  two situations where $(e_{\mathrm{IR}} /
  e_{\mathrm{vis}})_{\mathrm{lava}}$ is either larger (left) or smaller (right) than
  ($e_{\mathrm{IR}} / e_{\mathrm{vis}})_{\mathrm{rock}}$. The
  temperature curve is correctly retrieved in the first case while it is
  clearly not as accurate in the second case where a large fraction (30
  \%) of solutions are of catastrophic type.} \label{fig13}
  \end{figure*}

Let us now consider the case where the ratio $\epsilon_{\mathrm{IR}} /
\epsilon_{\mathrm{vis}}$ differs between the lava surface and the solid
ground. We compute the equilibrium temperature by making an additional
but physical assumption: the rocky part cannot be hotter than the lava
ocean at its bank. Indeed, if $(\epsilon_{\mathrm{IR}} /
\epsilon_{\mathrm{vis}})_{lava}$ > $(\epsilon_{\mathrm{IR}} /
\epsilon_{\mathrm{vis}})_{rock}$  when azimuth is larger than the one of the
lava ocean bank, the theoretical equilibrium temperature of the rocky
ground would be higher than the melting temperature: in that case, we
force it to be at the melting temperature. In doing so, we assume that there is a mix of solid and liquid as long as the rock
equilibrium temperature is not below the melting temperature. In 
Fig.~\ref{fig13}, we show two examples where the procedure of retrieving
the temperature curve is applied with $(\epsilon_{\mathrm{IR}} /
\epsilon_{\mathrm{vis}})_{lava}$ larger or smaller than
$(\epsilon_{\mathrm{IR}} / \epsilon_{\mathrm{vis}})_{\mathrm{rock}}$.
If the retrieved temperature curve is acceptable in the first case, obviously, the break at the bank can hardly be
restored by the fourth order polynomial expression in the second case, and the fraction of
catastrophic solutions is increased. At this stage, we do not try to go
further and use two polynomial expressions one for each
medium (lava or rock), but one can easily envision it when necessary.

If we retain that a modest SNR = 5 is sufficient to retrieve the
latitudinal temperature profile provided that the wavelength range
extends up to 4.5 $\mu m$, then we can estimate the time required to
obtain it by using the JWST Exposure Time Calculator in the case of
Corot-7b (K = 9.8) to be around 70 hours. This time could be split in
several periods of $\sim $ 3 hours to catch the planet close to to
the secondary transit where it offers the largest emitting surface
toward the observer. It can be anticipated that future missions (TESS,
PLATO) could detect a planet analog to Corot-7b around brighter stars and
 that the work will be easier in that case. For a G2V star
of magnitude K = 6 (V = 8) for instance, the  JWST ETC gives an integration time of 
9000 sec (2.5 h) that would be more easily acceptable given the
anticipated  competition for observing time.

\section{Conclusions}

In the context of an increasing number of detected very hot
super-Earths, such as Corot-7b or Kepler-10b,  we have explored whether
future observations with the instrument NIRSpec on JWST would allow us
to constrain their surface properties.  In particular, we have studied 
how the presence of a dense atmosphere or of a non synchronous rotation
of the planet could be inferred from multiwavelength light curves. We also studied to which accuracy one could retrieve the value of the surface Bond
albedo and the temperature distribution using the SED between 0.8 and
3.5 $\mu m$. 

We developed a model of the thermal behavior of the planet
surface by assuming that a lava lake is maintained around the substellar
point and by considering  various cases of heat penetration and albedo. This is done to estimate the time lag of the thermal emission when the planet
is not phase-locked and observed all along its orbit. We also modeled
the effect of a putative atmosphere at various pressures on the surface
temperature using an GCM  specifically developed for the study of
extrasolar planets. We  finally examined how the actual distribution of
temperature along a meridian could be retrieved from the observed SED by
assuming a polynomial distribution with latitude.  

Thanks to the broad range of wavelengths accessible with JWST and provided that  a signal to noise
ratio around 5 per resel is reached after a total exposure time of $
\sim$ 70 h,  we are generally able to constrain several of those
parameters with a fair accuracy : \begin{itemize} \item The
Bond albedo is retrieved to within $\pm 0.03$ in most cases. \item
Thanks to the asymmetry of the light curve due to the lag effect, the
rotation period of a non phase-locked planet is retrieved with an
accuracy of 3 hours   when the rotation is fast, or half the
orbital period. For a longer period, the accuracy is reduced.
\item The shortest period compatible with observation of a phase-locked
planet is in the range 30 - 800 h depending on the thickness
of the heat penetration layer. This means that any rotation period
shorter than this limit would be detectable. \item We should be able to
detect  the presence of a gray atmosphere, provided it is thick enough: with a pressure of one bar and an specific opacity higher than
$10^{-5} m^{-2} kg^{-1}$. \item The latitudinal temperature profile 
can be retrieved with an accuracy of 10 to 50 K depending on the signal
to noise ratio obtained with NIRSPEC (20 to 5 assumed). We note that the
risk of a totally wrong solution is not excluded but is at maximum of
5\% for a SNR of 5 and a wavelength range extending to 4.5 $\mu m$.
\end{itemize} Note that the four first results  would
still be valid if the exposure time was to be reduced to two 
complete orbital periods (i.e., 42h). 

These results show that we could test
the most current hypothesis of phase-locking and of an
atmosphere-free planet for planets similar to Corot-7b and confirm or reject the lava ocean model of L11 by using NIRSPEC-JWST spectra. A precise
determination  of the albedo and temperature distribution of the planet
surface appears also  doable, giving the  possibility to constrain more
precise models of the planet structure and properties. The possible
discovery of very hot exolanets around  stars  brighter than Corot-7 or
Kepler-10 would offer better opportunities to test those predictions: 
for instance, a more affordable integration time of only several hours 
would  be needed for a V = 8 solar-type star.

Of course, one cannot exclude that our model is too simplistic
and/or that the effects we did not envision could change some of the
conclusions on the feasibility.  For instance,
the hypothesis concerning a constant albedo, which is independent of wavelength, the nature and the phase (lava or solid) of the soil or, even more so, the hypothesis concerning a gray gas on the atmosphere may seriously affect the amplitude of the time lag effect we predict if the actual situation
is significantly different. On
the other hand, we consider the two results on the mean
abedo derivation and on the possibility to  retrieve
the latitudinal temperature profile to a large extent as rather solid. 

The access to the phase curves simultaneously in visible and near IR seems to be a promising source of information that will provide
useful constraints on the very hot super-Earths surfaces and in general on hot exoplanets. 

\bibliographystyle{aa}  % A&A bibliography style file (aa.bst)
\bibliography{samuel_biblio}

\end{document}